\documentclass{cpbtex}
\usepackage{xcolor}
\usepackage[most]{tcolorbox}
\usepackage{amsmath}
\allowdisplaybreaks

\begin{document}
\begin{CJK*}{GBK}{song}

\title{Maximal atom-photon entanglement in an N-type atomic system}

\author{Seyedeh Hamideh Kazemi, \ Nayyere Einali Saghavaz, and Mohammad Mahmoudi\\
{Department of Physics, University of Zanjan, University Blvd., 45371-38791, Zanjan, Iran}}

\date{\today}
\maketitle

\begin{abstract}

Atom-photon entanglement provides an essential resource for quantum communication and quantum computation. How to conveniently and efficiently achieve a maximal entanglement between atomic system and spontaneous emission field has been a challenging task. Here, we present a simple, yet we believe a powerful, method to generate entangled states between photons and an N-type atomic system. Beside the achievement of a nearly perfect entanglement, we also examine evidence for a link between entanglement and populations in dressed and bare states; It is found that a maximal entanglement can be established when populations in both dressed and bare states are spread over states. Moreover, the system would be disentangled in the absence of evenly distributed populations, another reason for further strengthen our claim that the physical origin of such entanglement is quantum correlation produced by distribution of the populations. We then discuss the dependence of the entanglement on Rabi frequency and detuning of the applied fields and demonstrate how an almost complete entanglement can be achieved for a judicious choice of these parameters. Note that entanglement is measured in semi-classical regime and by solving density matrix equations of motion and von-Neumann entropy. 

\end{abstract}



\section{Introduction}
Quantum entanglement, an interactive phenomenon in quantum systems composed of at least two sub-systems, has rightly been the focus of much work in recent years as a potential resource for quantum communication and information processing \cite{coffman}. Not only does this bizarre and beautiful aspect of quantum world play a leading role in foundations of quantum mechanics such as the violation of Bell's inequalities and the Einstein-Podolsky-Rosen (EPR) paradox \cite{EPR}, but it also provides a path for novel quantum technologies and quantum information applications including quantum coding \cite{8,8c}, quantum teleportation \cite{9,litele}, quantum positioning \cite{7a,Giovannetti}, quantum algorithm \cite{10} and quantum networking \cite{11}. 

Interaction between matter and light leads in general to diverse phenomena such as coherent population trapping (CPT) \cite{1,2,3}, electromagnetically induced transparency (EIT) \cite{eit1,eit2}, quantum correlation and entanglement \cite{en1,en2,en3,en4,en5,en6}. The latter phenomenon plays a vital role in quantum teleportation \cite{bennet,Bouwmeester}, quantum repeater \cite{dur}, long-distance quantum communication \cite{volz}, heralded generation of entanglement between quantum memories \cite{memory1,memory2} as well as a loophole-free Bell experiment \cite{simon}. So far, many different approaches for entangling light and matter have been proposed and used: One was presented in Ref.~\cite{duan} where the internal degree of freedom of the emitted photon can entangle with the spin-state of the atom/ion due to the spontaneous decay. Another method based on entanglement between coherently scattered photons and collective spin-excitations in atomic ensembles was suggested \cite{Matsukevich,chen}.

 One topic of particular interest in this context is the atom-photon entanglement where will be of special importance for the field of quantum communication and computation, since it allows to transmit quantum states over long distance \cite{andre}. It is apparent that both atom and photon can be regarded as a  carrier of storing and transmitting quantum information. While photons carry quantum information over arbitrary long distance with almost negligible decoherence, it is difficult to store them at fix location; The reverse is true for atoms. Up to know, atom-photon entanglement has been investigated in many previous experimental and theoretical studies. The work of Jauslin and co-workers, generating maximally entangled states of an atom and a cavity photon as well as two photons in  two spatially separate high-Q cavities, is mentioned as a prime example of this type of work \cite{amniat}. In another study, Fang \textit{et al.} investigated entanglement between a $\Lambda$-type three-level atomic system and its spontaneous emission field \cite{fang}. Recently, theoretical description on evolution of quantum entropy in a four-level atomic system has been investigated \cite{arzhang}. It has been also demonstrated by our group that atom-photon entanglement can be controlled by the relative phase of applied fields in a closed-loop atomic system \cite{kordi1}. 

On the other hand, the optical properties of a four-level N-type quantum system have been extensively studied \cite{25,26,30,27,28,ebrahimi}. For instance, resonance fluorescence, squeezing and absorption spectra of a laser-driven N-type system was investigated and it was demonstrated that quantum interference can induce two prominent and nearly transparent holes where the slope of the refractive index is very steep \cite{30}. Propagation of a weak probe field in a four-level N-type quantum system in the presence of spontaneously generated coherence (SGC) was theoretically investigated and found that group velocity of light pulse can be controlled by relative phase of applied fields \cite{ebrahimi}. Though the above-mentioned effects have been studied in the system, only a few studies have been performed on its quantum entangled properties and to the best of our knowledge, the maximal atom-photon entanglement in such system have not been achieved previously. 

In this study, we consider the interaction of an atom with three coherent laser fields and demonstrate an atom-photon entanglement in a four-level N-type quantum system. It is shown that in multi-photon resonance condition and in the presence of a superposition of populations in the dressed and the bare state, a nearly perfect entangled state between atomic system and its spontaneous emission field can be achieved. Accordingly, there would not be a mutual correlation between the system and the spontaneous emission fields, if there is no evenly distribution in the population. Finally, we discuss the robustness of the scheme against the laser parameters, the Rabi frequency and the detuning of the fields. Note that by solving density matrix equations of motion and by using the von-Neumann entropy, time evaluation of degree of entanglement (DEM) is employed to discuss the entanglement. This method would provide an efficient and convenient way to generate atom-photon entanglement by using only three laser fields, which may hold great promise for practical applications in quantum information processing.

\section{The Model} \label{sec:1}
The considered atomic system is composed by four-level atoms which are coupled by three laser fields, according to the scheme depicted in Fig.~1. Transitions  $ \vert 1 \rangle - \vert 3\rangle$, $\vert 2 \rangle - \vert 3\rangle$ and $\vert 1\rangle - \vert 4\rangle$ are driven by coherent fields $\Omega_{31}$, $\Omega_{32}$ and $\Omega_{41}$ with the corresponding carrier frequencies $\omega_{31}, \omega_{32}$ and $ \omega_{41}$. Moreover, laser field detuning are given by $\Delta_{31} = \omega_{31} - \omega_{31}, \Delta_{32} = \omega_{32} - \omega_{32}$ and $\Delta_{41} = \omega_{41} - \omega_{41}$. Noting that the general expression for a Rabi frequency is defined as $\Omega=(\bm \mu . \bm E)/{\hbar}$ with $\bm \mu$ and $\bm E$ being the atomic dipole moment of the corresponding transition and amplitude of the field, respectively. The spontaneous decay rates on the dipole-allowed transitions are denoted by $ \gamma_{41}$, $ \gamma_{42}$, $ \gamma_{32}$ and $ \gamma_{31}$. The Hamiltonian of the system in dipole and rotating wave approximation can be written as \cite{ebrahimi}
\begin{equation}
\mathrm{\hat{H}}= \hbar \Omega_{41} \exp(i \Delta_{41} t) \vert 4 \rangle \langle 1 \vert  + \hbar \Omega_{31} \exp(i \Delta_{31} t) \vert 3 \rangle \langle 1 \vert  + \hbar \Omega_{32} \exp(i \Delta_{32} t) \vert 3 \rangle \langle 2 \vert + \mathrm{H. c}.,
\end{equation}

\begin{figure}
\centering
\includegraphics[width=7cm]{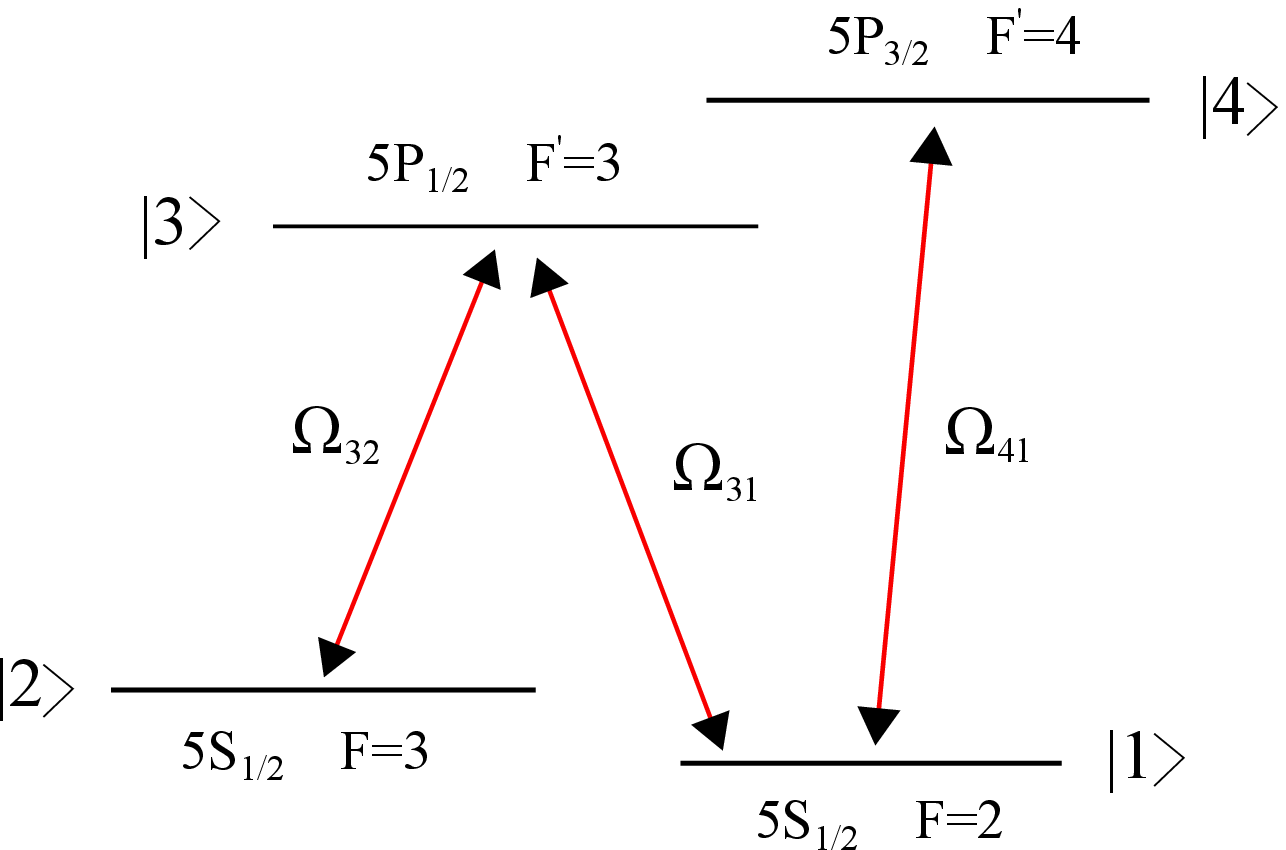}\\[5pt]
\parbox[c]{15.0cm}{\footnotesize{\bf Fig.~1.} The four-level N-type system of the transitions in $^{85} $Rb atom. }
\label{fig1}
\end{figure}

where H.c. corresponds to the Hermitian conjugate of the terms explicitly written in the Hamiltonian. Then, using the von-Neumann equation for density matrix, the equations of motion can be written as
\begin{subequations}
\begin{eqnarray}
\dot{\tilde{\rho}}_{11}&=& \gamma_{31}\tilde{\rho}_{33 }+ \gamma_{41}\tilde{\rho}_{44 } +i\vert \Omega_{31}\vert \tilde{\rho}_{31}- i\vert \Omega_{31}\vert \tilde{\rho}_{13} + i \Omega_{41} \tilde{\rho}_{41} -  i \Omega_{41} \tilde{\rho}_{14}, \\
\dot{\tilde{\rho}}_{22} &=&  \gamma_{32}\tilde{\rho}_{33 }+ \gamma_{42}\tilde{\rho}_{44 } +i\vert \Omega_{32}\vert \tilde{\rho}_{32} -i\vert \Omega_{32}\vert \tilde{\rho}_{23},\\
\dot{\tilde{\rho}}_{33} &=&   - (\gamma_{31} + \gamma_{32} ) \tilde{\rho}_{33 } +i\vert \Omega_{31}\vert \tilde{\rho}_{13} -i\vert \Omega_{31}\vert \tilde{\rho}_{31} + i \Omega_{32} \tilde{\rho}_{23} - i \Omega_{32} \tilde{\rho}_{32},\\ 
\dot{\tilde{\rho}}_{44}&=& - ( \gamma_{41} +\gamma_{42} ) \tilde{\rho}_{44 } +i\vert \Omega_{41}\vert \tilde{\rho}_{14} -i\vert \Omega_{41}\vert \tilde{\rho}_{41},\\ 
\dot{\tilde{\rho}}_{12}&=&  - i (\Delta_{31} - \Delta_{32}) \tilde{\rho}_{12 } + i\vert \Omega_{31}\vert \tilde{\rho}_{32} + i \vert \Omega_{41} \vert \tilde{\rho}_{42} - i \vert \Omega_{32} \vert \tilde{\rho}_{13}, \\ 
\dot{\tilde{\rho}}_{13}&=& -[\frac{ ( \gamma_{31} +\gamma_{32})}{2} +i\Delta_{31} ] \tilde{\rho}_{13}  -i \vert \Omega_{32}\vert \tilde{\rho}_{12} + i \vert \Omega_{31}\vert ( \tilde{\rho}_{33} -  \tilde{\rho}_{11})+ i \vert \Omega_{41} \vert \tilde{\rho}_{43}, \\
\dot{\tilde{\rho}}_{14}&=& -[\frac{ ( \gamma_{41} +\gamma_{42})}{2} +i\Delta_{41} ] \tilde{\rho}_{14} +i\vert \Omega_{31}\vert \tilde{\rho}_{34}  + i \vert \Omega_{41}\vert ( \tilde{\rho}_{44} -  \tilde{\rho}_{11}),\\
\dot{\tilde{\rho}}_{23}&=&   -[\frac{ ( \gamma_{31} +\gamma_{32})}{2} +i\Delta_{32} ] \tilde{\rho}_{23} -i \vert \Omega_{31}\vert \tilde{\rho}_{21}  + i \vert \Omega_{32}\vert ( \tilde{\rho}_{33} -  \tilde{\rho}_{22} ),\\
\dot{\tilde{\rho}}_{24}&=& -[\frac{ ( \gamma_{41} +\gamma_{42})}{2} +i (\Delta_{41} -\Delta_{31} +\Delta_{32}) ] \tilde{\rho}_{24} +i\vert \Omega_{32} \vert \tilde{\rho}_{34}  - i  \vert \Omega_{41} \vert \tilde{\rho}_{21},\\
\dot{\tilde{\rho}}_{34}&=&  -[\frac{ ( \gamma_{31} +\gamma_{32} +\gamma_{41} +\gamma_{42})}{2} + i (\Delta_{41} -\Delta_{31}) ] \tilde{\rho}_{34} +i\vert \Omega_{31} \vert \tilde{\rho}_{14}  + i \vert \Omega_{32}\vert  \tilde{\rho}_{24} \nonumber \\
&-&i \vert \Omega_{41} \vert \tilde{\rho}_{31}.
\end{eqnarray}
\label{eq2}
\end{subequations}
The remaining equations follow from the constraints $\tilde{\rho}_{ij}=\tilde{\rho}^{*}_{ji}$ and $\sum _{i} \tilde{\rho}_{ii}=1 $. We have also defined rotating frame as
\begin{eqnarray}
\tilde{\rho_{12}} &=& \rho_{12} e^{i [ ( \Delta_{32} - \Delta_{31} ) t + ( \phi_{32} - \phi_{31}) ]} , \quad \tilde{\rho_{13}} = \rho_{13} e^{i  ( \Delta_{31} t + \phi_{32})}, \,  \tilde{\rho_{14}} = \rho_{14} e^{ - i \Delta_{41}t }\nonumber \\
 \tilde{\rho_{23}} &=& \rho_{23} e^{ - i ( \Delta_{32} t - \phi_{32})}, \quad \quad \quad  \quad  \quad \, \,
 \tilde{\rho_{24}} =\rho_{24} e^{ - i [ ( \Delta_{32}+ \Delta_{41} -  \Delta_{31} ) t + ( \phi_{31} - \phi_{32}) ]} \nonumber \\
\tilde{\rho_{34}} &=& \rho_{34} e^{ - i [ ( \Delta_{41} - \Delta_{31} ) t +  \phi_{31} ]}.
\end{eqnarray}

In what follows, we proceed to quantify entanglement. It is well known that the physical essence of entanglement lies in the fact of quantum correlations between individual parts of a composite system that have interacted once in the past but are no longer interacting. In the case of a bipartite quantum system, quantification of entanglement is simple and straightforward; The reduced density matrix of a subsystem (A or B) $\rho_{A}$ or $\rho_{B}$ is pure if the state of the composite system ($\rho_{AB}$) is pure as well as separable. Keeping in mind $\rho_{A}= \mathrm{tr}_{B} \, \rho_{AB}$ and $\rho_{B}= \mathrm{tr}_{A} \, \rho_{AB}$ where the partial trace has been taken over one subsystem, either A or B. If the composite system is entangled, $\rho_{A}$ and $\rho_{B}$ are mixed even though $\rho_{AB}$ may be pure. So, the degree of entanglement of a pure state for such system is associated with the degree of mixing of its reduced states. 

Generally, von-Neumann entropy is used to quantify the degree of mixing of a reduced state $\rho_{A}$ or $\rho_{B}$ which has its origins in seminal work of Vedral \textit{et al.} \cite{vedral}. It is also imperative to point out that von-Neumann entropy entropy possesses appropriate properties to be a good measure of entanglement in the case of a bipartite state \cite{akbari,knight}. It is additive in such a way that the entanglement of two copy of a same state is double of the entanglement of a single copy. Moreover, it has asymptotic continuity so that small changes in a state can lead to the changes in its entropy \cite{chandra}. The von-Neumann entropy is given by \cite{vedral2}

\begin{equation}
S = -\mathrm{tr}(\rho_A \,\ln \rho_A)= -\mathrm{tr}(\rho_B \, \ln \rho_B) ,
\end{equation}
where, $S$ is zero in the case of a disentangled pure joint state. 

In this study, we deal with two subsystem A and F (say the atom and field), so by assuming the reduced density operator of the atoms (spontaneous emission) as the first (the second) subsystem $\rho_{A(F)}$ and the density operator of the pure state for two subsystems by $\rho_{AF}$, the partial von-Neumann entropy is derived as $ S_{A(F)} = -\mathrm{tr}(\rho_{A(F)} \, \ln \rho_{A(F)})$. We also consider that the atom and the field are initially in a disentangled pure state, meaning that all of the atoms are initially in just one state (ground state) so that the overall system can be a pure state and the entropy of one subsystem can be used to measure its degree of entanglement with other subsystem. As notably pointed out by Araki and Lieb \cite{akbari}, for a bipartite quantum system composed of two subsystems at any time t, these entropies satisfy the triangle inequalities
\begin{equation}
\vert S_{A}(t)- S_{F}(t)  \vert \leq S_{AF}(t) \leq \vert S_{A}(t)+ S_{F}(t) \vert.
\label{eq10}
\end{equation}
Here, $S_{AF}$ is the total entropy of the composite system. Because of being the trace is invariant under similarity transformation in which the atomic density matrix is diagonal, the DEM for atom-field entanglement is defined by
\begin{equation}
\mathrm{DEM}(t)= S_A(t)=S_F(t)= -\sum\limits^{4}_{i=1} \lambda_i \, \ln \lambda_i,
\label{eq11}
\end{equation}
with $\lambda_i$ being as the eigenvalue of the reduced density matrix.

\section{Results and discussions}

Before presenting numerical results, we point out some important considerations: As a realistic example, we consider hyperfine energy levels of $ ^{85}$Rb; Two upper levels, i.e., $\vert 3\rangle$ and $\vert 4\rangle$ correspond, respectively, to the sublevels $ \vert 5 P_{1/2}, F^{'}=3\rangle$ and $ \vert 5 P_{3/2}, F^{'}=4 \rangle$, while $\vert 5 S_{1/2}, F=2\rangle$ and $\vert 5 S_{1/2}, F=3\rangle$ are chosen to be the lower states ($\vert 1\rangle$ and $\vert 2\rangle $), respectively. We also assume $\Delta_{31}=\Delta_{32}=\Delta_{41}=\Delta$ and $\gamma_{31}=\gamma_{32}=\gamma_{41}=\gamma_{42}=\gamma$.

In what follows, we numerically calculate the degree of entanglement between the atom and its spontaneous emission fields by using equations~\ref{eq2} and~\ref{eq11} in which the time evaluation of the DEM is employed to discuss the entanglement. We assume that the atomic system is initially in the ground state ($\vert 1 \rangle$) and the relevant parameters are scaled with $\gamma$. In Fig.~2, we plot the dynamical behavior of DEM for three different detunings, $\Delta =0$ (solid line), $\Delta =2 \gamma$ (dashed line), $\Delta =4 \gamma$ (dotted line) and for $\Omega_{31}=\Omega_{32}= \Omega_{41}=5 \gamma $. An investigation on Fig.~2 shows that how the DEM depends on this parameter and increases form 0.7 at $\Delta =4 \gamma$ to 1.36 at $\Delta=0$.

\begin{figure}
\centering
\includegraphics[width=9cm]{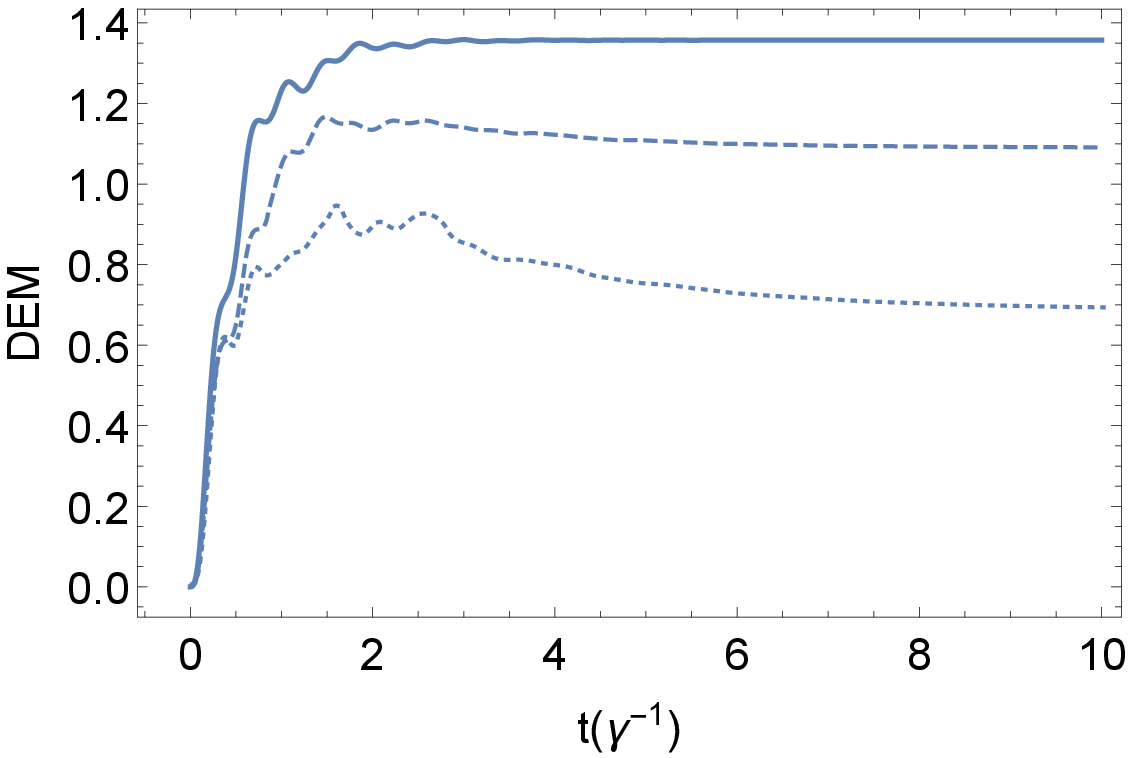}\\[5pt]
\parbox[c]{15.0cm}{\footnotesize{\bf Fig.~2.} Time evolution of DEM for $\Omega_{31}=\Omega_{32}= \Omega_{41}=5 \gamma $ and different detunings, $\Delta =0$ (solid line), $\Delta =2 \gamma$ (dashed line) and $\Delta =4 \gamma$ (dotted line). }
\label{fig2}
\end{figure}

We now introduce dressed states generated by applied fields in which the physics of the DEM can be explained via population distribution of the dressed states. The calculated dressed states for $\Omega_{31}=\Omega_{32}= \Omega_{3}$ and in the multi-photon resonance condition, $\Delta=0$, are as follows

\begin{eqnarray}
\vert \mathrm{D}_1 \rangle &=& - \dfrac{\sqrt{ B- A }}{\sqrt{2} \Omega_{41}} \vert 1 \rangle + \dfrac{\sqrt{B- A  } (\Omega_{41}^{2}+A)}{2 \sqrt{2} \Omega_{3}^{2} \Omega_{41}} \vert 2 \rangle  -  \dfrac{A+C }{2 \Omega_{3} \Omega_{41}} \vert 3 \rangle + \vert 4 \rangle, \nonumber\\
\vert \mathrm{D}_2 \rangle &=&  \dfrac{\sqrt{B- A }}{\sqrt{2} \Omega_{41}} \vert 1 \rangle - \dfrac{\sqrt{B- A  } (\Omega_{41}^{2}+A)}{2 \sqrt{2} \Omega_{3}^{2} \Omega_{41}} \vert 2 \rangle  - \dfrac{A+C }{2 \Omega_{3} \Omega_{41}} \vert 3 \rangle + \vert 4 \rangle, \nonumber\\
\vert \mathrm{D}_3 \rangle &=& - \dfrac{\sqrt{ A+B }}{\sqrt{2} \Omega_{41}} \vert 1 \rangle + \dfrac{\sqrt{A+B  } (\Omega_{41}^{2}- A)}{2 \sqrt{2} \Omega_{3}^{2} \Omega_{41}} \vert 2 \rangle  - \dfrac{C-A }{2 \Omega_{3} \Omega_{41}} \vert 3 \rangle + \vert 4 \rangle, \nonumber\\
\vert \mathrm{D}_4 \rangle &=&  \dfrac{\sqrt{A +B}}{\sqrt{2} \Omega_{41}} \vert 1 \rangle - \dfrac{\sqrt{ A+B  } ( \Omega_{41}^{2} -A)}{2 \sqrt{2} \Omega_{3}^{2} \Omega_{41}} \vert 2 \rangle  - \dfrac{C-A}{2 \Omega_{3} \Omega_{41}} \vert 3 \rangle + \vert 4 \rangle,  \nonumber\\
\end{eqnarray}
where we define $A=\sqrt{4 \Omega_{3}^{4} + \Omega_{41}^{4}}$, $B= 2 \Omega_{3}^{2} + \Omega_{41}^{2}$ and $C= 2 \Omega_{3}^{2} - \Omega_{41}^{2}$ in order to obtain a compact form of the above expressions.

\begin{figure}
\centering
\includegraphics[width=14cm]{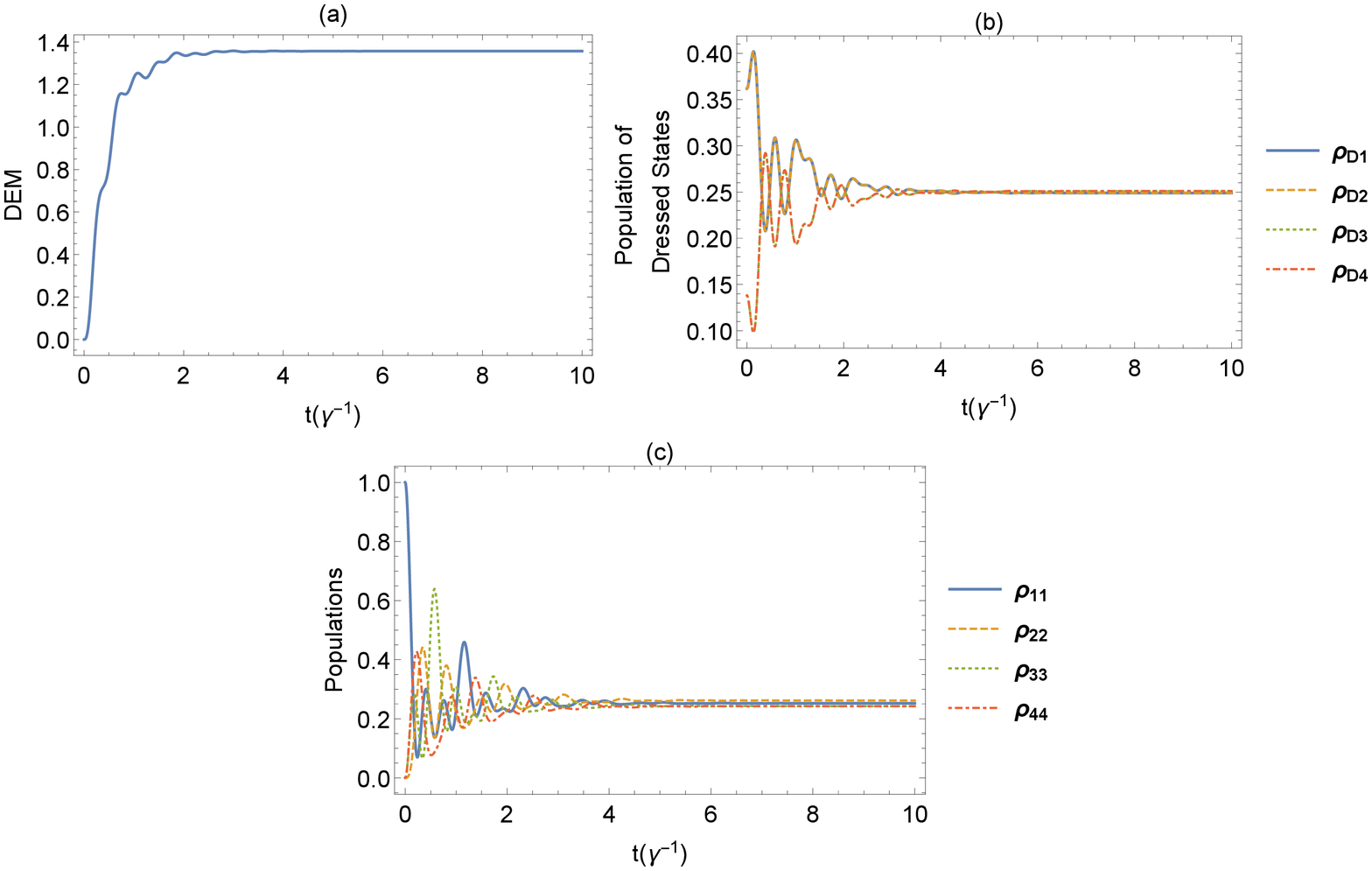}\\[5pt]
\parbox[c]{15.0cm}{\footnotesize{\bf Fig.~3.} (a) Time evolution of the DEM, (b) time evolution of population in the dressed states and (c) population in the bare states in multi-photon resonance condition, $\Delta=0$. Other parameters are the same as in Fig.~2.}
\label{fig3}
\end{figure}

In the following part of the article, we present our analysis discussing dynamics of the system, i.e., DEM, population in dressed state as well as that in bare states. Unless otherwise stated explicitly, we restrict our consideration to multi-photon resonance condition. As can be seen from Fig.~3, these curves share a main feature: Reaching a value at time of about $5\gamma$ and then fixing at the value. First, let us say some words about the time-dependence of the DEM which is shown in Fig.~3(a). This degree of the entanglement between atom and photon has major increase during the process due to distribution of dressed states, which will be discussed below; Starting from zero, it quickly rises to its maximum with a value of 1.36 and then will be fixed at such value, indicating that the atom and its spontaneous emission fields are strongly entangled. We then proceed to accumulate evidence for a link between entanglement and population in dressed and bare states. To understand the physical mechanism of this behavior, especially sudden increase of the DEM, we take a glance at time evolution of population in the dressed states (Fig.~3(b)). It can clearly be observed that the population in dressed state is spread over four states. If we compare this plot with the curve of Fig.~3(a), we clearly see that the majority of time-dependent features of the DEM seem to be inherited from the dynamics of the population, in particularly the timing of the increases synchronize with that of the DEM. Thereby, it is legitimized to conclude that by evenly distributing the population over the dressed state, bipartite entanglement is established. Noting that this result is not surprising from physical intuition, because this distribution of the dressed states would lead to quantum correlating. A similar behavior is also observed in population of bare states which will be distributed rather evenly between the states as time goes on (Fig.~3(c)). We can easily observe that main features of this curve are very similar to those of the dressed states: After rising and falling several times during the process, the population stays constant after passing time of $5\gamma$, around a value 0.25.  

\begin{figure}
\centering
\includegraphics[width=14cm]{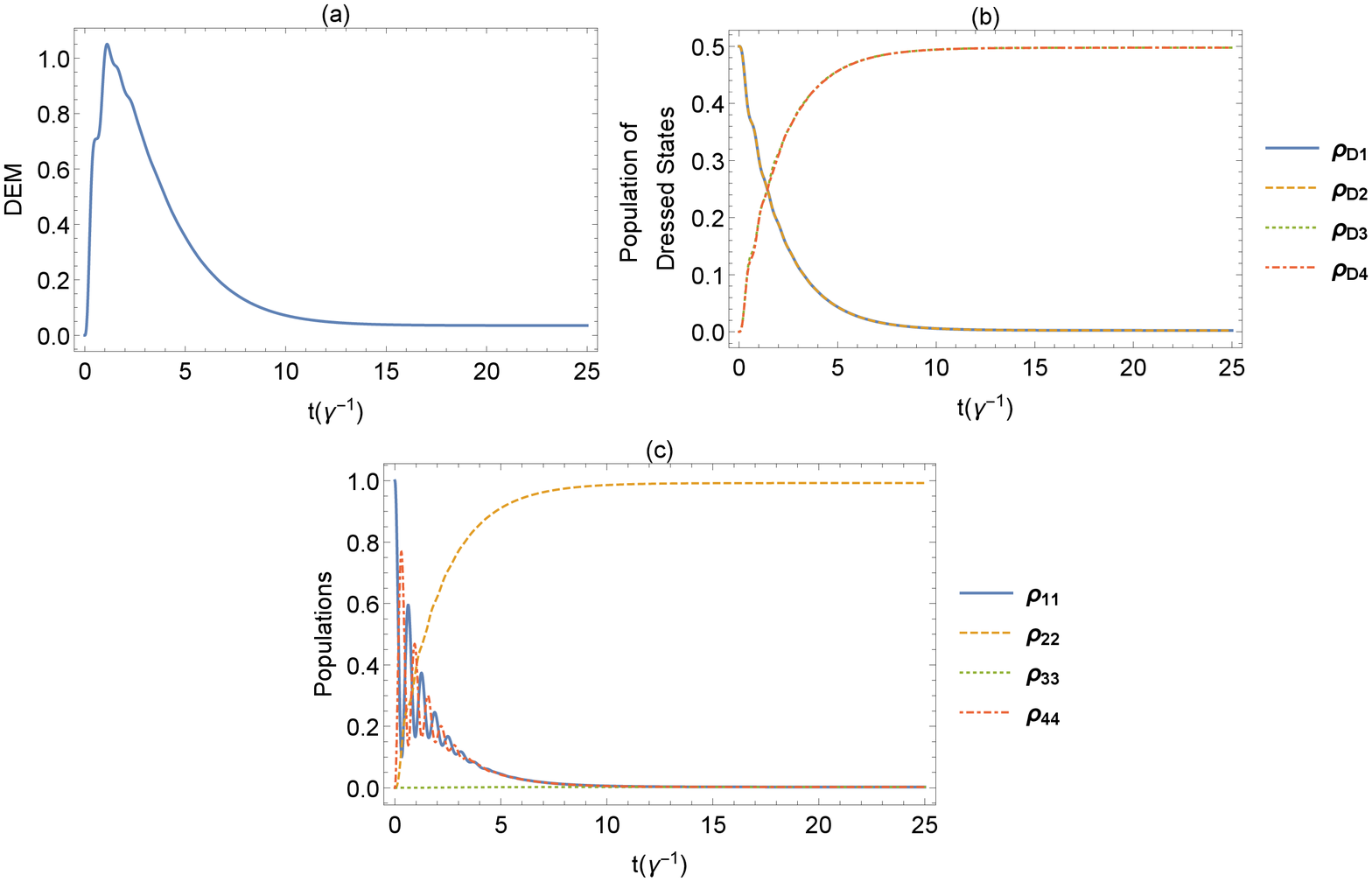}\\[5pt]
\parbox[c]{15.0cm}{\footnotesize{\bf Fig.~4.} (a) Time evolution of the DEM, (b) time evolution of population in the dressed states and (c) population in the bare states for $\Omega_{31}=\Omega_{32}= 0.05 \gamma $ and $\Omega_{41}=5 \gamma$. Other parameters are the same as in Fig.~3.  }
\label{fig4}
\end{figure}

To restate our argument that the physical mechanism underlying the entanglement resides in the distribution of the population, in somewhat different and perhaps more efficient ways, we present Fig.~4 where the dynamical behaviors of the DEM, population of different dressed and bare states are plotted. We stated in previous paragraph that the maximal entanglement can be found if the population in the dressed states will be distributed evenly between the states, therefore we expect that the entanglement should be significantly decreased in the absence of such distribution. Fig.~4 shows more clearly the striking remark that was already present; Only in the presence of evenly distributed population, a maximal atom-photon entanglement is achieved. 
It is clearly shown by this figure that the DEM approaches zero at long time when population in the dressed and bare states are not spread over  four states, instead transferring to either one or two states. From Fig.~4(a) and (b), we can see that no correlation and so no entanglement exists because the population in the dressed states are 0 or 0.5. Obviously, the complete transfer of population of the bare states will eliminate entanglement of the system. 

\begin{figure}
\centering
\includegraphics[width=9cm]{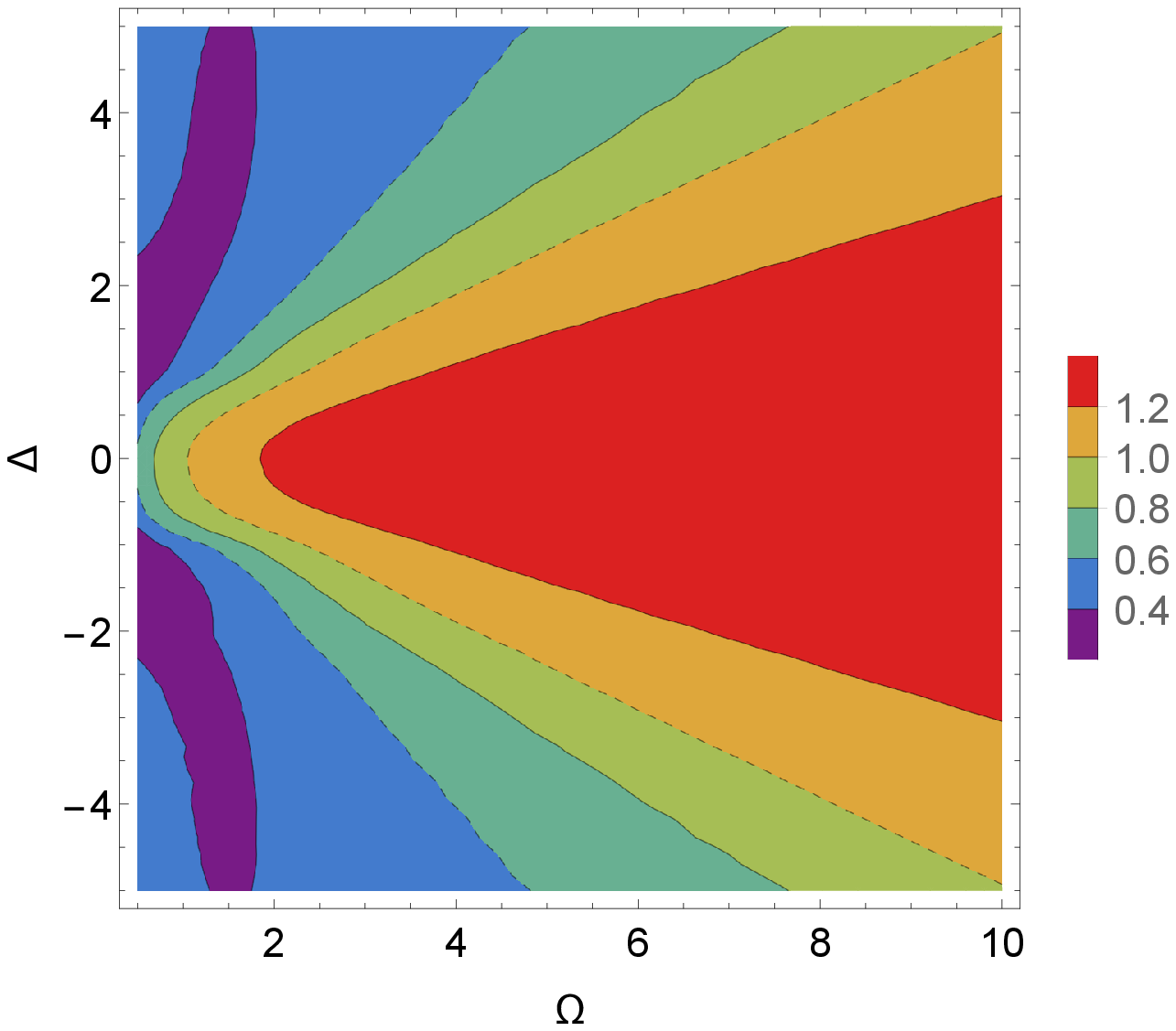}\\[5pt]
\parbox[c]{15.0cm}{\footnotesize{\bf Fig.~5.}  The contour plots of the steady-state DEM as a function of the Rabi frequency $\Omega$ and the detuning $\Delta$ of the fields. }
\label{fig5}
\end{figure}

Now we turn our attention to evaluate the robustness of the suggested scheme against parameters of applied fields. In Fig.~5, we present the contour plots of the steady-state DEM i.e., the value that remains constant as time evolves, as a function of the Rabi frequency and the detuning of the fields in which for simplicity, we assume equal Rabi frequencies $\Omega_{31}=\Omega_{32}=\Omega_{41}=\Omega$. A careful inspection of the figure reveals that one can be able to achieve an almost complete entanglement for a judicious choice of these parameters. For instance, entanglement with value more than 1.2 takes place with Rabi frequency distributing in the range of 2$\gamma$ to 10$\gamma$ and for detuning in the range of $-0.5 \gamma$ to $0.5 \gamma$. Also, we note that in a large range of variation in these parameters, one can obtain entanglement with the DEM more than 1. We will further explore the DEM, with also putting forward convincing argument concerning mechanism underlying the entanglement by presenting result from analytical solutions of the density matrix.

Here, it is worth comparing our scheme with a previous one involving time evaluation of quantum entropy in a four-level N-type atomic system \cite{arzhang}. In addition to obvious difference of coupling the atom to the laser fields, there are also differences in the condition of entanglement; They found that the degree of entanglement can be increased by quantum interference induced by spontaneous emission, i.e., SGC, while rigorous condition of non-orthogonal dipole moments and also near degenerated states are barely met in real atomic systems and no experiments has been conducted yet in atoms to observe the SGC directly. Moreover, they assumed that atomic system is initially in the superposition of upper levels. So, from the viewpoint of practical applications, the entanglement produced in our study seems to be more promising. Apart from the technical difficulties, their steady-state atom entropy seems unsatisfactory as they achieved a maximal entanglement of only 1, less than both maximum value of entanglement and our results. 

Then, we proceed to derive an analytical expression for the coherence and the eigenvalues in
order to support our argument that physical origin underlying the entanglement resides in the distribution of the population. In the case of $\Delta=0$ and for $\Omega_{31}=\Omega_{32}=\Omega_{41}=\Omega$, expression of the steady-state coherence, which can be obtained from equations~2 for vanishing time derivatives, yields

\begin{equation}
\tilde{{\rho}}_{32}= \tilde{{\rho}}_{41}= \tilde{{\rho}}_{42}=D=\dfrac{I \gamma \Omega^{3}}{4 \Omega^{4} + 2 \gamma^{2} \Omega^{2}}. 
\end{equation}
For $\Omega < 5 \gamma$, we have $\tilde{{\rho}}_{ii} \approx 0.25$ and can write these expressions for the eigenvalues 
\begin{eqnarray}
\lambda_{1} &=& \dfrac{1}{4} (1-2\sqrt{2} \sqrt{ (3 -\sqrt{5} )\,  D^2} ), \, \lambda_{2} = \dfrac{1}{4} (1+2\sqrt{2} \sqrt{ (3 -\sqrt{5} )\,  D^2} ), \nonumber\\
\lambda_{3} &=& \dfrac{1}{4} (1-2\sqrt{2} \sqrt{ (3 +\sqrt{5} )\,  D^2} ), \, \lambda_{4} = \dfrac{1}{4} (1+2\sqrt{2} \sqrt{ (3 +\sqrt{5} )\,  D^2} ).\nonumber\\
\end{eqnarray}

\begin{figure}
\centering
\includegraphics[width=9cm]{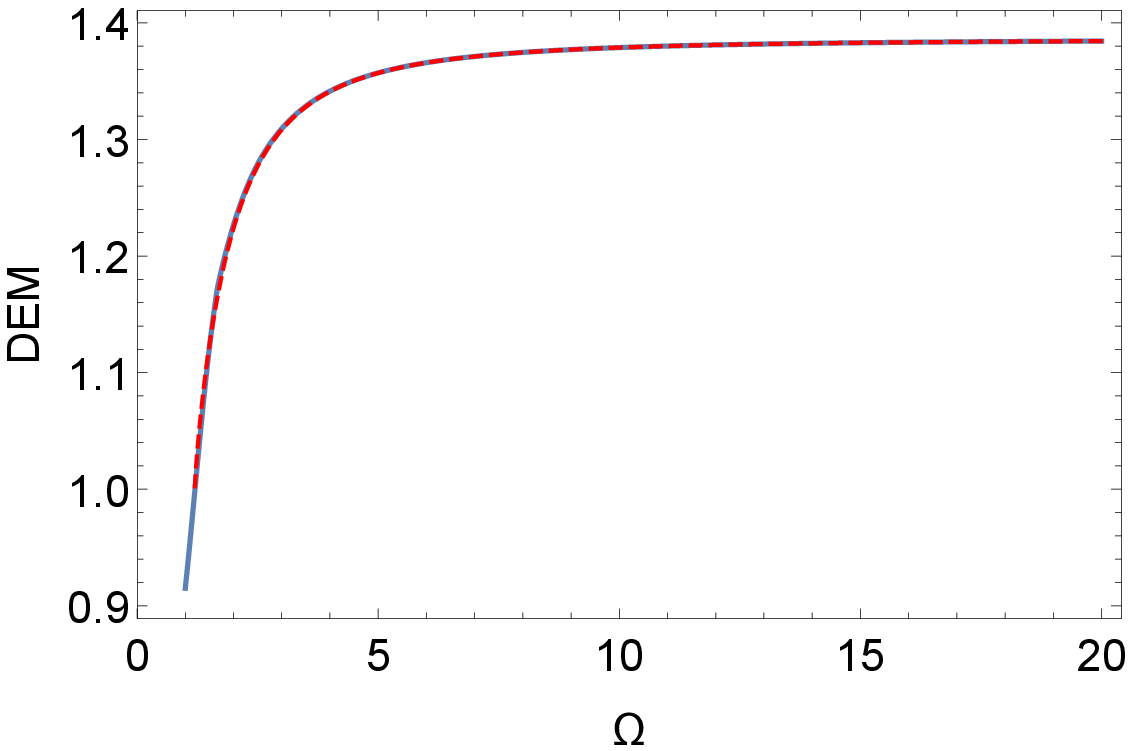}\\[5pt]
\parbox[c]{15.0cm}{\footnotesize{\bf Fig.~6.} The calculated steady-state DEM from numerical solutions of the density matrix (solid line) and its analytical solutions (dashed line) as a function of the Rabi frequency $\Omega$. Other parameters are the same as in Fig.~3. }
\label{fig6}
\end{figure}

The analytical results for steady-state DEM are even more interesting if we compare them to exact one, so we display steady-state DEM calculated from numerical solutions of the density matrix (solid line) and the analytical solutions (dashed line) via the Rabi frequency in Fig.~6. Despite deviations appeared for small Rabi frequency- mainly due to the fact that the population of the bare states are not exactly 0.25-, these curves exhibit a very good agreement and match in an acceptable range of the Rabi frequency. Another remark which seems rather important is that the entanglement is found to be enhanced by increasing the Rabi frequency, reaching value of 1.38 or nearly the maximum value of entanglement (Ln(4) $\simeq $ 1.386). Moreover, the above results further confirm our argument that there is a link between between the DEM and the distribution of the population; For the Rabi frequencies larger than $5 \gamma$, we have $\tilde{{\rho}}_{ii} \approx 0.25$, i.e., population of the bare states is distributed between the states and consequently an almost complete entanglement can be achieved (see the results of Fig.~6 and also equation~9).

Finally, we envision that experimental demonstration of our scheme can be implemented with laser-cooled alkali atoms. As an example, we can consider hyperfine energy levels of $^{85} $Rb (see Fig.~1 for more details). We also note that using a collisional blockade mechanism, we can ensure only single atoms can be found \cite{reymond}. With regard to photon, it is spontaneously emitted from the atom which is excited to a state having multiple decay channels. A magnetic field can provide a quantization axis in such a way that only photons emitted along the axis are collected. After loading the atom into a trap, initialing an internal atomic qubit state and then exciting it for the spontaneous emission of the photon, entangling is started and after detecting the emitted photon, the entanglement can be verified through the analysis of the photon and state detection of the atom. In this respect, it is worthwhile to mention a work of Weinfurter \textit{et al.} in which entanglement between a single trapped atom and a single photon at a wavelength of 780 nm- suitable for low-loss communication over the long distances- is observed and a single atom state analysis is introduced in order to increase fidelity, without any further atomic-state manipulation \cite{volz}. These methods, accompanied by quantum gates, can form the basis for quantum communication and photonic quantum communication channel.

\section{Conclusions}

Entanglement of a four-level N-type atomic system that is coupled to three laser fields has been traced, discussed and a simple method to generate entangled state between photons and this system is proposed. A measurement of the entanglement between the atom and its spontaneous emission fields, through the DEM, has been presented. It has been found that in the presence of the superposition of population in the dressed and the bare state, a nearly perfect entanglement could be established under  the multi-photon resonance condition. Moreover, there would not be a mutual correlation between the system and the spontaneous emission fields and therefore no entanglement would be established, if there is no evenly distribution in the population. These findings along with our analytical solutions emphasized strong relation between the population distribution and the entanglement. Further, an investigation into whether the scheme is robust against parameters of applied fields has been carried out. We must reiterate the importance of the fact the maximal entanglement can be always achieved for a large range of the parameters of the fields, while the entanglement in the similar study is 1, under favorable circumstances and only for limited parameters and limited applicability. We vision that our approach opens doors for additional investigations in numerous quantum systems and facilitate the generation of atom-photon entanglement.

\end{CJK*}  
\end{document}